%% file: main.tex
\documentclass[adoi4paper, UKenglish, cleveref, autoref, thm-restate, pdfa]{lipics-v2021}

\input{macro}
\DeclareMathOperator{\ctx}{ctx}
\DeclareMathOperator{\type}{type}

\pdfoutput=1
\hideLIPIcs
\nolinenumbers

\title{Mechanized HOL Reasoning in Set Theory}

\author{Simon Guilloud}{EPFL, Laboratory for Automated Reasoning and Analysis, Switzerland \and \url{https://people.epfl.ch/simon.guilloud} }{}{https://orcid.org/0000-0001-8179-7549}{}

\author{Sankalp Gambhir}{EPFL, Laboratory for Automated Reasoning and Analysis, Switzerland \and \url{https://people.epfl.ch/sankalp.gambhir} }{}{https://orcid.org/0000-0001-5994-1081}{}

\author{Andrea Gilot}{EPFL, Laboratory for Automated Reasoning and Analysis, Switzerland \and \url{https://people.epfl.ch/andrea.gilot} }{}{https://orcid.org/0009-0006-4463-9414}{}

\author{Viktor Kun\v{c}ak}{EPFL, Laboratory for Automated Reasoning and Analysis, Switzerland \and \url{https://lara.epfl.ch/~kuncak/} }{}{https://orcid.org/0000-0001-7044-9522}{}

\authorrunning{S. Guilloud, S. Gambhir, A. Gilot and V. Kun\v{c}ak} 

\Copyright{Simon Guilloud, Sankalp Gambhir, Andrea Gilot and Viktor Kun\v{c}ak} 

\ccsdesc[100]{Theory of computation~Logic and verification} 

\keywords{Proof assistant, First Order Logic, Set Theory, Higher Order Logic} 

\date{2024}

\begin{document}
\maketitle

\begin{abstract}
    We present a mechanized embedding of higher-order logic (HOL) and algebraic data types (ADT) into first-order logic with ZFC axioms. We implement this in the Lisa proof assistant for schematic first-order logic and its library based on axiomatic set theory.
    HOL proof steps are implemented as proof producing tactics in Lisa, and the types are interpreted as sets, with function (or arrow) types coinciding with set-theoretic function spaces.
    The embedded HOL proofs, as opposed to being a layer over the existing proofs, are interoperable with the existing library. This yields a form of soft type system supporting top-level polymorphism and ADTs over set theory, and offer tools to reason about functions in set theory.
\end{abstract}

\section{Introduction}

The interactions and combinations of higher-order logic (HOL) with set theory in the context of proof systems have been a long-standing topic of study (e.g. \cite{brownHigherOrderTarskiGrothendieck2019, 
gordonSetTheoryHigher1996, krauss2010mechanized, Brown2006Combining, agerholm1995Experiments}). Set theory (in particular ZF and ZFC) is the prototypical and oldest formalized foundation of mathematics, but it does not naturally admit a concept of ``typed'' expressions, which are widely used in informal mathematics, for guiding automated proof search, and in programming languages. HOL on the other hand naturally supports typed expressions, type checking and reasoning for simply typed lambda calculus with top level polymorphism.

The first goal of the present work is to study a syntactic embedding of HOL into first-order set theory. It is well known that the Zermelo-Franekel axioms (ZF) imply the existence of a set that is a model of higher-order logic, where types are interpreted as sets, type judgement as set membership and $\lambda$-terms as set-theoretic functions
%
%
\cite{gordonSetTheoryHigher1996}. However, the mere fact that models of set theory contain a model of HOL (that is, a semantic embedding) is of little use by itself in practice. To be able to write the syntax and simulate the features of HOL, we need a \emph{syntactic} embedding that transforms expressions in HOL into terms and formulas of first order set theory.

Some proof assistants have explored mixing features of set theory and HOL in various ways in their foundations, such as Egal~\cite{brownEgalManual2014}, Isabelle/HOLZF~\cite{obuaPartizanGamesIsabelle2006} or  ProofPeer~\cite{Obus2015Type}.
A translation of statements and proofs from HOL to set theory has been done for some systems, for example between Isabelle/HOL and Isabelle/ZF \cite{krauss2010mechanized}, between Isabelle/HOL and Isabelle/Mizar \cite{Kaliszyk2023Combining, brownHigherOrderTarskiGrothendieck2019} or between HOL Light and Metamath \cite{carneiro2016conversion}. However, both Isabelle/ZF and Metamath, as well as other systems using higher-order Tarski-Grothendieck \cite{brownHigherOrderTarskiGrothendieck2019}, the Hilbert $\epsilon$ operator (as suggested in \cite{gordonSetTheoryHigher1996}) or built-in notations for replacement and comprehension allow, in some form, writing \emph{terms} containing \emph{bound variables}. 
This is impossible in syntactically strict first-order logic as described in mathematical text books and in first-order automated theorem provers and proof assistants, where only the universal and existential quantifiers, $\forall$ and $\exists$, can bind variables. This makes it impossible to naturally express arbitrary $\lambda$-terms of the form $\lambda x. t$ as standalone first-order terms.

Nonetheless, we show that it is possible to embed higher-order logic and $\lambda$-terms in first-order set theory without these constructs using a notion of \textit{contexts}, i.e. formulas that gives local assumptions about terms. We study how this embedding impacts decision procedures for type checking and simulating the proof steps of HOL and implement the embedding in the Lisa proof assistant \cite{guilloudLISAModernProof2023}, whose foundations are built on first-order set theory. We obtain from this a form of soft type system over set theory, and support for reasoning about functions with HOL-like proofs steps in Lisa. Specifically, we implement the proof steps of HOL Light \cite{harrisonHOLLightOverview2009}, for the simplicity of its foundations. Incidentally, we hoped this implementation would allow automatic import of theorems from the HOL Light library. However, while this embedding this works well in practice for human-written proofs, which typically don't contain a many high towers of nested $\lambda$-abstractions, our initial tests suggest that the embedding has too much overhead in the size of the proofs for the translation of large proofs whose basic building blocks are such $\lambda$-abstractions to be of practical use.

In the second part of this paper, with the same motivation of simulating features from type systems into set theory, we describe how Algebraic Data Types (ADTs) can be encoded into set theory. ADTs are types defined inductively by their constructors, one of the simplest examples is that of singly-linked lists of integers, given by $\text{List} = \text{Nil} \mid \text{Cons}(\text{head}: \mathbb{Z}, \text{ tail}: \text{List})$. ADTs and their generalizations are essential constructs in type theory based proof assistants (such as Coq \cite{the_coq_development_team_2023_8161141} and Lean \cite{demouralean4}) and in functional programming languages. Given the description of an ADT in terms of the type signature of its constructors, we show how to define the set corresponding to the type and functions representing the constructors, deriving the desired theorems about induction and injectivity. We show that expressions using ADTs and their constructors can be type checked by the same procedure as expressions from higher-order logic embedded in Lisa. Finally, we extend these results to polymorphic ADTs.

\subsection{Contributions}

The contribution of this paper is to present a practical embedding of simply typed lambda calculus with polymorphism, of the proof steps of HOL Light and of ADTs into classical ZFC within first-order logic, and its implementation in the Lisa proof assistant.
\begin{itemize}
    \item We describe how to embed simply typed lambda calculus (and in particular lambda abstractions, which cannot be syntactically expressed in first-order logic) into set theory. Our approach is based on maintaining a context of local definitions, expressing the desired properties about the subterms of $\lambda$-terms of the form $\lambda x. t$. If $t$ contains variables other than $x$, i.e. free variables, we need to encode the closure of $\lambda x.t$ instead, similar to the compilation of programs containing nested function declaration.
    \item We explain how this encoding allows simulating proof steps and type checking from HOL by producing the corresponding proofs in set theory.
    \item We implement this embedding and the proof-producing tactics in the Lisa proof assistant\footnote{Disclaimer: as of March 2024, the library of results from Lisa is still under development. As such, some intermediate results regarding set-theoretic function space and the recursion theorem are still ``assumed''.}, allowing reasoning about set-theoretic functions using HOL proof steps. \item We try to use this embedding to import (parts of) the library of HOL Light, but obtain a negative result because the simulation is too complex and indirect.
    \item We describe how ADTs can be automatically defined in ZF set theory and how to obtain their key recursive properties derived from the recursion theorem in ZF set theory. We mechanize this system in Lisa, making ADTs and their constructors fully compatible with implemented HOL tactics.
\end{itemize}


In the present work we picked HOL Light as our reference system for HOL, but with some additional work, the results can be translated to other proof assistants in the HOL-family of proof assistants, such as HOL4 \cite{slindBriefOverviewHOL42008}, Isabelle/HOL \cite{nipkowIsabelleHOLProof2002}, or Candle \cite{abrahamssonCandleVerifiedImplementation2022}.

Similarly, our target system was Lisa, but none of the results are specific to Lisa, and they transfer to any system using axiomatic set theory over first-order logic, though those are not quite as common, such as Mizar \cite{naumowiczBriefOverviewMizar2009}. Our implementation can be found in {\small\url{https://github.com/epfl-lara/lisa/tree/itp2024-archive}}.

\newcommand{\B}{ \mathcal B}
\newcommand{\ind}{ i}
\newcommand{\Vl}{ V^\lambda}
\newcommand{\Tl}{ T^\lambda}
\newcommand{\F}{ F}
\renewcommand{\P}{ P}
\newcommand{\vl}{ v^\lambda}
\newcommand{\tl}{ t^\lambda}
\newcommand{\ve}{ v}
\newcommand{\te}{ t}
\newcommand{\app}{\mbox{app}}
\newcommand{\isFunction}{\mbox{isFunction}}
\newcommand{\isTotal}{\mbox{isTotal}}
\newcommand{\isFunctional}{\mbox{isFunctional}}

\newcommand{\fostl}{FOST$^\lambda${ }}

\section{Preliminaries}

\input{preliminaries}

\section{From HOL Formulas to First-Order Set Theory Formulas}
\input{HOL}

\section{Formalizing Algebraic Data Types}
\input{ADT}

\section{Importing Proofs from HOL Light}
\input{hol-import}

\section{Conclusion}
We have demonstrated how to embed HOL into conventional first-order logic axiomatization of set theory. Our translation maintains local definitions (at the level of the sequent) of the closure of abstraction terms. We showed that this encoding allows simulating all the core proof steps of higher-order logic. We mechanized this encoding in Lisa, and obtained an interface and tactics for reasoning with typed expressions and set-theoretic functions.
We then demonstrated how (possibly polymorphic) ADTs can be mechanized in first-order set theory, and that their representation is compatible with the tactics and type checking we developed for HOL functions.

We also considered alternative encodings of lambda terms. 
Instead of defining lambdas at the level of the whole sequent, we could place the definition right after the first predicate symbol. In particular, definitions of lambdas become nested, instead of being independent. On one hand, this means that we do not have to compute closures. On the other hand, the defining property of a lambda would often be deep in a formula, and its use would require deconstructing and reconstructing the formula to use the context.
Alternatively, we could use an embedding of $\lambda$-terms based on combinators from combinatory logic~\cite{barendregtLambdaCalculusTypes2013}. We did not use fixed combinators such as SKI due to growth in formula size; in the future we may explore the use of parametric combinator families. 
%

While the results we obtain are of practical use and we expect them to become part of the standard Lisa release, the encoding is somewhat complicated and even if most of the machinery can be hidden, it may confuse new users. There is a significant overhead in the size of formulas and the simulation of proofs. This overhead can in the worst case be linear in the size of the formulas, and even if those do not tend to grow indefinitely, a large constant factor may be less than ideal in practice. This also prevented us from importing a larger number of theorems from the HOL Light library.
 The syntactic restrictions on terms of FOL is the main source of complexity in the translation. For Lisa, this suggests considering extensions of FOL with terms that refer to formulas, such as the definite description operator $\iota x. P$, denoting an individual that is uniquely characterized by the predicate $P$. 


{
\bibliographystyle{plainurl}
\raggedright
\bibliography{biblio.bib,vkuncak.bib, sguilloudZotero.bib}
}

\newpage
\appendix
\section{Appendix}

\subsection{Deduction Rules and axioms for FOL and ZF}
\begin{figure}[h!]
    \begin{center}
        \begin{tabular}{l l}
            \AxiomC{}
            \RightLabel{\text { Hypothesis}}
            \UnaryInfC{$\Gamma, \phi \vdash \phi, \Delta$}
            \DisplayProof
            &
            \AxiomC{$\Gamma \vdash \phi, \Delta$}
            \AxiomC{$\Sigma, \phi \vdash \Pi$}
            \RightLabel{\text{ Cut}}
            \BinaryInfC{$\Gamma, \Sigma \vdash \Delta, \Pi$}
            \DisplayProof
            \\[4ex]

            \AxiomC{$\Gamma \vdash \Delta$}
            \RightLabel{\text { LeftWeakening} }
            \UnaryInfC{$\Gamma, \phi \vdash \Delta$}
            \DisplayProof
            & 
            \AxiomC{$\Gamma \vdash \Delta$}
            \RightLabel{\text { RightWeakening} }
            \UnaryInfC{$\Gamma \vdash \phi, \Delta$}
            \DisplayProof
            \\[4ex]

            \AxiomC{$\Gamma, \phi, \psi \vdash \Delta$}
            \RightLabel{\text { LeftAnd}}
            \UnaryInfC{$\Gamma, \phi \land \psi \vdash \Delta$}
            \DisplayProof 
            &
            \AxiomC{$\Gamma \vdash \phi, \Delta$}
            \AxiomC{$\Sigma \vdash \psi, \Pi$}
            \RightLabel{\text{ RightAnd}}
            \BinaryInfC{$\Gamma, \Sigma \vdash \phi \land \psi,  \Delta, \Pi$}
            \DisplayProof
            \\[4ex]

            \AxiomC{$\Gamma, \phi \vdash \Delta$}
            \AxiomC{$\Sigma, \psi \vdash \Pi$}
            \RightLabel{\text{ LeftOr}}
            \BinaryInfC{$\Gamma, \Sigma, \phi\lor \psi \vdash \Delta, \Pi$}
            \DisplayProof 
            &
            \AxiomC{$\Gamma \vdash \phi, \psi \Delta$}
            \RightLabel{\text{ RightOr}}
            \UnaryInfC{$\Gamma \vdash \phi \lor \psi,  \Delta$}
            \DisplayProof
            \\[4ex]

            \AxiomC{$\Gamma \vdash \phi, \Delta$}
            \RightLabel{\text { LeftNot}}
            \UnaryInfC{$\Gamma, \neg \phi \vdash \Delta$}
            \DisplayProof 
            &
            \AxiomC{$\Gamma, \phi \vdash \Delta$}
            \RightLabel{\text{ RightNot}}
            \UnaryInfC{$\Gamma \vdash \neg \phi ,  \Delta$}
            \DisplayProof
            \\[4ex]

            \AxiomC{$\Gamma, \phi_x[t := x] \vdash \Delta$}
            \RightLabel{\text { LeftForall}}
            \UnaryInfC{$\Gamma, \forall x. \phi_x  \vdash \Delta$}
            \DisplayProof 
            &
            \AxiomC{$\Gamma \vdash \phi_x, \Delta$}
            \RightLabel{\text { RightForall}}
            \UnaryInfC{$\Gamma \vdash \forall x. \phi_x,  \Delta$}
            \DisplayProof
            \\[4ex]

            \AxiomC{$\Gamma, \phi_x \vdash \Delta$}
            \RightLabel{\text { LeftExists}}
            \UnaryInfC{$\Gamma, \exists x. \phi_x \vdash \Delta$}
            \DisplayProof 
            &
            \AxiomC{$\Gamma \vdash \phi_x[x := t], \Delta$}
            \RightLabel{\text { RightExists}}
            \UnaryInfC{$\Gamma \vdash \exists x. \phi_x,  \Delta$}
            \DisplayProof
            \\[4ex]

            \AxiomC{$\Gamma, \phi[x := t] \vdash \Delta$}
            \RightLabel{\text{ LeftSubstEq}}
            \UnaryInfC{$\Gamma, t=u, \phi[x := u] \vdash \Delta$}
            \DisplayProof 
            &
            \AxiomC{$\Gamma \vdash \phi[x := t], \Delta$}
            \RightLabel{\text{ RightSubstEq}}
            \UnaryInfC{$\Gamma, t = u \vdash \phi[x := u], \Delta$}
            \DisplayProof
            \\[4ex]
            
            \AxiomC{}
            \RightLabel{\text { Refl}}
            \UnaryInfC{$\vdash t = t$}
            \DisplayProof 
            &
            \AxiomC{$\Gamma \vdash \Delta$}
            \RightLabel{\text{ Inst}}
            \UnaryInfC{$\Gamma[\vec x := \vec t] \vdash \Delta[\vec x := \vec t]$}
            \DisplayProof 
        \end{tabular}
    \end{center}
\caption{Deduction rules for first order logic with equality.}
\label{fig:rulesFOLappendix}
    \begin{center}
    \small
        \begin{tabular}{ll}
            \AxiomC{}
            \RightLabel{\text { EmptySet}}
            \UnaryInfC{$\forall x. x \notin \emptyset$}
            \DisplayProof
            &
            \AxiomC{}
            \RightLabel{\text { Extensionality}}
            \UnaryInfC{$(\forall z. z \in x \iff z \in y) \iff (x = y)$}
            \DisplayProof
            \\[4ex]

            \AxiomC{}
            \RightLabel{\text { Subset}}
            \UnaryInfC{$x\subset y \iff \forall z. z \in x \implies z \in y$}
            \DisplayProof
            &
            \AxiomC{}
            \RightLabel{\text { Pair}}
            \UnaryInfC{$(z \in \lbrace x, y\rbrace) \iff ((x = z) \lor (y = z))$}
            \DisplayProof
            \\[4ex]

            \AxiomC{}
            \RightLabel{\text { Union}}
            \UnaryInfC{$(z \in \operatorname{U}(x)) \iff (\exists y. (y \in x) \land (z \in y))$}
            \DisplayProof
            &
            \AxiomC{}
            \RightLabel{\text { Powerset}}
            \UnaryInfC{$(x \in \operatorname{\mathcal{P}}(y)) \iff (x \subset y)$}
            \DisplayProof
            \\[4ex]
            
            \AxiomC{}
            \RightLabel{\text { Foundation}}
            \UnaryInfC{$\forall x. (x \neq \emptyset) \implies (\exists y. (y \in x) \land (\forall z. z \in x))$}
            \DisplayProof
            &
            \AxiomC{}
            \RightLabel{\text { Comprehension}}
            \UnaryInfC{$\exists y. \forall x. x \in y \iff (x \in z \land \phi(x))$}
            \DisplayProof
            \\[4ex]

            \multicolumn{2}{l}{
                \AxiomC{}
                \RightLabel{\text { Infinity}}
                \UnaryInfC{$\exists x. \emptyset \in x \land (\forall y. y \in x \implies \operatorname{U}(\lbrace y, \lbrace y, y \rbrace \rbrace) \in x)$}
                \DisplayProof
            }\\[4ex]

            \multicolumn{2}{l}{
                \AxiomC{}
                \RightLabel{\text { Replacement}}
                \UnaryInfC{$
                    \forall x. (x \in a) \implies \forall y, z.  (\psi(x, y) \land \psi(x, y)) \implies y = z \implies 
                    (\exists b. \forall y. (y \in B) \implies (\exists x. (x \in a) \land \psi(x, y)))
                $}
                \DisplayProof
            }\\[4ex]

        \end{tabular}
    \end{center}

  \caption{Axioms for Zermelo-Fraenkel set theory.}
  \label{fig:axiomszf}
\vspace{-3em}
\end{figure}

\newpage
\subsection{ZF Transfinite Recursion Schema}

\begin{thm}[Transfinite Recursion]
    Let $F$ be a class function and $\alpha$ an ordinal. There exists a unique function $f$ with domain $\alpha$ such that 
    \begin{center}
        $\forall \beta \in \alpha.\ f(\beta) = F(f \arrowvert_{\beta})$
    \end{center}
\end{thm}
\begin{proof}
    See \cite[p.~22]{jech2003set}, \cite[p.~25]{kunenSetTheoryIntroduction1983}. 
\end{proof}
In Lisa this translates into

\begin{lstlisting}[language=scala]
Theorem( ordinal(a) |- 
    existsOne(f, functionalOver(f, a) /\ 
        forall(b, in(b, a) ==> 
            (app(f, b) === F(restrictedFunction(f, b)))
        )
    )
)
\end{lstlisting}

\end{document}

%% file: macro.tex
\usepackage{url}
\usepackage{graphicx} 
\usepackage[utf8]{inputenc}
\usepackage{amsmath}
\usepackage{amsfonts}
\usepackage{aliascnt}
\usepackage{amsthm}
\usepackage{bussproofs}
\usepackage{dirtytalk}
\usepackage{float}
\usepackage{xspace}
\usepackage{hyperref}
\usepackage{xifthen}
\usepackage[nomap]{FiraMono}
\usepackage{tikz}
\usepackage{stmaryrd}
\usepackage{framed}
\usetikzlibrary{positioning}
\usepackage[ruled,vlined,linesnumbered]{algorithm2e}
    \SetAlFnt{\fontsize{8.5}{10}\selectfont}
    \SetArgSty{textup}
    \SetProgSty{textup}
    \SetFuncArgSty{textup}
    \SetFuncSty{textit}
    \SetKwProg{procedure}{def}{}{}
    \SetKwProg{datastructure}{Datastructure}{}{}
    \SetKwSwitch{Switch}{Case}{Other}{match}{:}{case}{otherwise}{endcase}{endsw}
    
    \SetKw{KwCont}{continue}
    \SetKw{KwBreak}{break}

\SetKwComment{Comment}{/* }{ */}

\theoremstyle{definition}

\newtheorem{thm}{Theorem}[section]

\newtheorem{lem}[thm]{Lemma}

\newtheorem{ex}[thm]{Example}

\newtheorem{cor}[thm]{Corollary}

\newtheorem{defin}[thm]{Definition}

\newcommand{\OL}{\mathit{O{\kern-0.1em}L}}

\newcommand{\FOLm}{\mathit{F{\kern-0.1em}({\kern-0.1em}O{\kern-0.1em}L{\kern-0.1em})\textsuperscript{2}}}

\newcommand{\ssimw}[1]{{{\kern-0.05em}/{\kern-0.08em}{#1}}}

\newcommand{\sime}{\sim_{{\kern-0.1em}E}}
\newcommand{\siml}{\sim_{{\kern-0.1em}L}}

\newcommand{\eqclass}[2]{{[#1]}_{#2}}
\newcommand{\eqclasssim}[2]{\eqclass{#1}{\sim_{{\kern-0.1em}#2}}}

\newcommand{\simon}[1]{}
\newcommand{\viktor}[1]{}
\newcommand{\sankalp}[1]{}

\usepackage[nomap]{FiraMono} 

\definecolor{green}{rgb}{0, 0.6, 0}

\definecolor{comments}{RGB}{80,0,110}

\lstdefinelanguage{scala}{
    alsoletter={@,=,>},
    keywordstyle = {\color{blue}},
    keywordstyle = [2]{\color{blue}},
    commentstyle = \color{comments},
    morekeywords = [2]{abstract, case, class, def, do, Input, Output, then,
        else, extends, false, free, if, implicit, match,
        object, true, val, var, while, sealed, or,
        for, dependent, null, type, with, try, catch, finally,
        import, final, return, new, override, this, trait,
        private, public, protected, package, throw},
    sensitive = true, 
    numbers=left,
    stepnumber=1,
    morecomment = [l]{//},
    morecomment = [s]{/*}{*/},
    morestring = [b]",  
    otherkeywords = {;,<<,>>,++},
    mathescape = true,
    escapeinside = {{*@}{@*}},
    literate = {
        {λ}{\(\lambda\)}{1}
    },
}

\lstdefinelanguage{lisa}{
    alsoletter={@,=,>},
    keywordstyle = {\color{blue}},
    keywordstyle = [2]{\color{blue}},
    keywordstyle = [3]{\color{green}},
    keywordstyle = [4]{\color{teal}},
    commentstyle = \color{comments},
    morekeywords = [2]{abstract, case, class, def, do, Input, Output, then,
        else, extends, false, free, if, implicit, match,
        object, true, val, var, while, sealed, or,
        for, dependent, null, type, with, try, catch, finally,
        import, final, return, new, override, this, trait,
        private, public, protected, package, throw},
    morekeywords = [3]{have, andThen, thenHave, Theorem, by, DEF, The, Lemma, subproof, assume},
    sensitive = true, 
    numbers=left,
    stepnumber=1,
    morecomment = [l]{//},
    morecomment = [s]{/*}{*/},
    morestring = [b]",  
    otherkeywords = {;,<<,>>,++},
}

\newcommand{\interp}[1]{\left\llbracket #1 \right\rrbracket}

\newcommand{\embed}[1]{\llparenthesis #1 \rrparenthesis}

\newcommand{\vdashhol}{\vdash}
\newcommand{\vdashfost}{\vdash}

\newcommand{\ALPHACONV}{\lstinline{ALPHA_CONV} }
\newcommand{\EQMP}{\lstinline{EQ_MP} }
\newcommand{\uEQMP}{\lstinline{_EQ_MP} }
\newcommand{\ALPHAEQUIVALENCE}{\lstinline{ALPHA_EQUIVALENCE} }
\newcommand{\uTRANS}{\lstinline{_TRANS} }
\newcommand{\TRANS}{\lstinline{TRANS} }
\newcommand{\BETA}{\lstinline{BETA} }
\newcommand{\ETA}{\lstinline{ETA} }
\newcommand{\INST}{\lstinline{INST} }
\newcommand{\INSTTYPE}{\lstinline{INSTTYPE} }
\newcommand{\ABS}{\lstinline{ABS} }
\newcommand{\CLEAN}{\lstinline{CLEAN} }
\newcommand{\REFL}{\lstinline{REFL} }
\newcommand{\ASSUME}{\lstinline{ASSUME} }
\newcommand{\MKCOMB}{\lstinline{MK_COMB} }
\newcommand{\DEDUCT}{\lstinline{DEDUCT_ANTISYM_RULE} }

\newcommand{\consttag}{\text{tag}}

%% file: preliminaries.tex
There exist several different variations of higher-order logic (HOL). We consider it as it is defined in HOL Light. Its language is the simply typed lambda calculus with top-level polymorphism (or, the Hindley-Milner type system). Each variable in an HOL term is associated with a type, which may contain type variables. The deduction rules of HOL Light are described in \autoref{fig:rulesHOL}\footnote{HOL Light also admits a choice function and an infinity axiom later in the library development, which are justified in ZFC by the choice axiom and infinity axiom. These two additional axioms are largely tangential to the concerns of the present work and hence outside the scope. While \ETA is also formally an axiom in HOL Light, we consider it because set-theoretic functions are naturally extensional, and to handle alpha-equivalence in \autoref{subsec:SyntacticEmbedding}.}. We denote by $\B$ the type of Booleans containing two elements, by $\ind$ the infinite type of individuals, and by $=^\lambda$ the built-in function representing equality.

\begin{figure}[ht!]
    \begin{center}
        \begin{tabular}{ll}
            \AxiomC{}
            \RightLabel{\text { \REFL}}
            \UnaryInfC{$\vdashhol t =_A t$}
            \DisplayProof
            &
            \AxiomC{$\Gamma \vdashhol s =_A t$}
            \AxiomC{$\Delta \vdashhol t =_A u$}
            \RightLabel{\text { \TRANS}}
            \BinaryInfC{$\Gamma, \Delta \vdashhol s =_A u$}
            \DisplayProof
            \\[4ex]

            \AxiomC{$\Gamma \vdashhol s =_A t$}
            \AxiomC{$\Delta \vdashhol u =_A v$}
            \RightLabel{\text { \MKCOMB}}
            \BinaryInfC{$\Gamma, \Delta \vdashhol s(u) =_A t(v)$}
            \DisplayProof
            &
            \AxiomC{$\Gamma \vdashhol s =_B t$}
            \RightLabel{\text { \ABS}}
            \UnaryInfC{$\Gamma \vdashhol (\lambda x:A. s) =_{A \to B} \lambda(x:A.t)$}
            \DisplayProof
            \\[3ex]

            \AxiomC{}
            \RightLabel{\text { \BETA}}
            \UnaryInfC{$\vdashhol \lambda (x:A. t)x =_B t$}
            \DisplayProof
            &
            \AxiomC{}
            \RightLabel{\text { \ASSUME}}
            \UnaryInfC{$p \vdashhol p$}
            \DisplayProof
            \\[3ex]

            \AxiomC{$\Gamma \vdashhol p =_{\mathcal{B}} q$}
            \AxiomC{$\Delta \vdashhol p$}
            \RightLabel{\text { \EQMP}}
            \BinaryInfC{$\Gamma, \Delta \vdashhol q$}
            \DisplayProof
            &
            \AxiomC{$\Gamma, q \vdashhol p$}
            \AxiomC{$\Delta, p \vdashhol q$}
            \RightLabel{\text { \DEDUCT}}
            \BinaryInfC{$\Gamma, \Delta \vdashhol p =_{\mathcal{B}} q$}
            \DisplayProof
            \\[3ex]

            \AxiomC{$\Gamma \vdashhol p$}
            \RightLabel{\text{\INST}}
            \UnaryInfC{$\Gamma[\vec x := \vec t] \vdashhol p[\vec x := \vec t]$}
            \DisplayProof 
            &
            \AxiomC{$\Gamma \vdashhol p$}
            \RightLabel{\text{ \INSTTYPE}}
            \UnaryInfC{$\Gamma[\vec X := \vec A] \vdashhol p[\vec X := \vec A]$}
            \DisplayProof \\
            
            \AxiomC{}
            \RightLabel{\text {\ETA}}
            \UnaryInfC{$\vdashhol (\lambda x. t x) =_A t$}
            \DisplayProof
        \end{tabular}
    \end{center}
    \caption{Deduction rules for higher-order logic as implemented in HOL Light.}
    \label{fig:rulesHOL}
    
\end{figure}

Note that the Infinity axiom asserts that $\ind$ is infinite. It requires definition of logical connectives in HOL, which we do not explain in the present work.

We also assume familiarity with the syntax and deduction rules of first-order logic (see, for example, \cite{mendelsonIntroductionMathematicalLogic1987}). We assume Sequent Calculus \cite{gentzenUntersuchungenUberLogische1935} as our proof system, which is used by our proof assistant Lisa, but the results transfer to other proof systems for first-order logic (FOL). We call \textit{first-order set theory} (FOST) the axiomatic system of ZFC~\cite{jech2003set, kunenSetTheoryIntroduction1983} in first order logic. This is also the foundation of the list Lisa proof system~\cite{guilloudLISAModernProof2023} in which we implement our result.

In both HOL (see \cite{arthanHOLConstantDefinition2014}) and FOL, the language can be extended conservatively using the concept of \textit{extension by definition}, as described (for example) in \cite[Section 2.10]{kunenSetTheoryIntroduction1983} and \cite{guilloudLISAReferenceManual2023}.

\begin{thm}[Extension by Definition for First Order Logic]
\label{thm:extFOL}
    Let $K$ be a first order theory (for example, FOST), and $\phi$ a formula with free variables $y, x_1,...,x_n$. 
    Suppose $\vdash_K \forall x_1,...,x_n \exists ! y. \phi$ and let the theory $K'$ be $K$ with the addition of a function symbol $f$ of arity $n$ and the axiom $\forall y, y = f(x_1,...,x_n) \iff \phi$. Then $K'$ is fully conservative over $K'$
\end{thm}

\subsection{Set-Theoretic Semantics of HOL}
To motivate our translation from HOL to set theory, we review classical set-theoretic semantics of HOL. This allows us to focus first on the semantics of functions and types, without having  to worry about if a certain set is expressible, efficiently or at all, as a term in FOST. We interpret types as sets and HOL functions as total set theoretic functions.

As is usual in set theoretic foundations, we identify a function $f : A \to B$ with its graph, that is, as a subset of $A \times B$ such that for every element $x$ of $A$, there exists exactly one element $y \in B$ where $(x,y) \in f$. We write $\isFunction(f, A, B)$ to denote that the set $f$ is a total and functional relation (or simply, a function) from $A$ to $B$.

We define an operator $\app$ such that, for all $f$ such that $\isFunction(f, A, B)$ and for all $x\in A$, $\app(f, x) = y$ iff $(x, y) \in f$.

\begin{defin}[Set theoretic universe]
\label{def:setUniverse}
We use the following concepts to give a classical 
semantics to HOL.
\begin{itemize}
    \item Fix $U$ to be a set that is a universe of Zermelo set theory, i.e., a containing an infinite set, and closed under powersets, unions, and subsets defined by set comprehension (separation axiom). Consequently, $U$ is closed under Cartesian products. For example, we can take $U$ to be the set $V_{\omega+\omega}$ of the cummulative (von Neumann) hierarchy \cite[Chapter 6]{jech2003set}.
    \item Let $\app$ and $\isFunction$ be as above.
    \item For $A, B \in U$, let 
        $A \Rightarrow B$ denote $\lbrace r \in \mathcal P (A \times B) \mid \isFunctional(r, A, B) \rbrace$
    \item Let $\mathbb N$ be the set of natural numbers.
    \item Let $\bot = \emptyset$, $\top = \lbrace \emptyset\rbrace$ and $\mathbb B = \lbrace \bot, \top\rbrace$   
    \item For $A \in U$, let $E(A) =$
    $$ \lbrace (x, f) \in (A \times (A \Rightarrow \mathbb B)) \mid f = \lbrace (y, b) \in (A \times \mathbb B) \mid  (x = y \rightarrow b = \top) \land (x \neq y \rightarrow b = \bot) \rbrace\rbrace$$
    Note that for all $A$, $E(A) \in (A \Rightarrow (A \Rightarrow \mathbb B))$

\end{itemize} 
\end{defin}

\begin{defin}[Semantics of HOL]
    \label{defin:shallowEmbedding}
    An assignment $\alpha: (\Vl \cup \Tl) \to U$ is a function such that for all $x:A \in \vl$, $\alpha(x:A) \in \alpha(A)$.
    We define an interpretation of HOL terms with respect to an assignment:
    $$
    \begin{array}{lll}
        \interp{X}_\alpha & = & \alpha(X) \\
        \interp{\Tl_1 \to \Tl_2}_\alpha & = & \interp{\Tl_1}_\alpha  \Rightarrow
        \interp{\Tl_1 }_\alpha \\
        \interp{\mathcal B}_\alpha & = & \mathbb B\\
        \interp{i}_\alpha & = & \mathbb N \\
        \interp{x:A}_\alpha & = & \alpha(x) \\
        \interp{=^\lambda : A \to A \to \mathcal B} & = & E(\interp{A}_\alpha) \\ 
        \interp{(f: A \to B)(t: A) : B}_\alpha & = & \app(\interp{f: A \to B }_\alpha, \interp{t:A}_\alpha)\\
        \interp{ (\lambda x: A.\ t:B): A \to B}_\alpha & = & 
            \lbrace (y, z) \in (\interp{A}_\alpha \times \interp{B}_\alpha) \mid z = \interp{t}_{\alpha[x \mapsto y]} \rbrace 
    \end{array}
    $$
\end{defin}

\begin{defin}[Syntactic and Semantic truth]\ \\
For any FOST sequent $s = (l_1,...,l_n) \vdash (r_1,...,r_n)$, we write:
\begin{itemize}
    \item $\vdashfost s$ if $s$ is provable in FOST
    \item $U \models s$ if $U$ satisfies $(l_1 = \top \land .. \land l_n = \top) \implies (r_1 = \top \lor .. \lor r_n = \top)$ in the usual sense of first order models.
\end{itemize}
For any HOL sequent $s = (l_1,...,l_n) \vdash r$, we write:
\begin{itemize}
    \item $\vdashhol s$ if $s$ is provable in HOL
    \item $U \models s$ if, for every assignment $\alpha$, $(\interp{l_1}_\alpha = \top \land .. \land \interp{l_n}_\alpha = \top) \implies \interp{r}_\alpha = \top$ holds in $U$
\end{itemize}
\end{defin}

\begin{thm}
    For any term $s : A \in \tl$ and assignment $\alpha$, $\interp{s:A}_\alpha \in \interp{A}_\alpha$
\end{thm}
\begin{proof}
    By induction on the structure of $t$.
\end{proof}
We can show that all rules  of HOL from \autoref{fig:rulesHOL} hold in ZFC, giving the following theorem:
\begin{thm}
    For any assignment $\alpha$ and HOL terms $s_1:\B,...s_n:\B$ and $t:\B \in \tl$,
    $$ \text{if }(s_1,...,s_n \vdashhol t \text{ and } \forall i. \ \interp{s_i: \mathcal B}_\alpha = \top) \text{ then } \interp{t: \mathcal B}_\alpha = \top$$
    
\end{thm}

While the above argument shows that HOL has an interpretation into first order set theory, it does not immediately give us a mechanical translation from an HOL proof system to proofs in mechanized set theory. In particular, note that in \autoref{defin:shallowEmbedding}, the right-hand side of the lambda case cannot be expressed in the syntax of first order logic. It does not either tell us if and how we can automate the translation of an HOL proof into a FOST proof, or the production of proofs of a statement $t \in A$ that would correspond to type checking. However, note also that the embedding is shallow, in the sense that HOL functions are interpreted as usual set theoretic functions and types of functions as sets of set theoretic functions.

%% file: HOL.tex
We wish to define a translation $\embed{\cdot}$ from HOL sequents to FOL sequents such that if an HOL sequent $s$ is provable in HOL then $\embed{s}$ is provable in FOST. Technically, a trivial such embedding would map all sequents to the trivially true sequent. We cannot require that the embedding maps unprovable sequents of HOL to unprovable sequents of FOST, because FOST is strictly more powerful and can prove additional statements. But, we can require that the embedding does not map semantically false statements to provable sequents. This means, for every sequent $s$ of HOL:
\begin{enumerate}
    \item $\vdashhol s\ \implies\ \vdashfost \embed{s}$
    \item $\vdashfost \embed{s}\ \implies\ U \models s$
\end{enumerate}

Moreover, for the embedding to be of practical use in theorem proving and for import of proofs, we would like the embedding to be as natural as possible, so that we ideally have an embedding $\embed{\cdot}: \tl \to \te$, i.e. from terms of HOL to terms of FOST, such that, for every assignment $\alpha$, we have 
$
\interp{\embed{s:A}}_\alpha = \interp{s:A}_\alpha
$.
Unfortunately, the syntax of FOST terms does not support $\lambda$-abstractions. Of course, the set we denote by
$$
\lbrace (y, z) \in (\interp{A}_\alpha \times \interp{B}_\alpha) \mid z = \interp{t}_{\alpha[x \mapsto y]} \rbrace 
$$
is guaranteed to exist by the Comprehension axiom, but the above expression is not a term in first-order logic. In particular, any variable that appears in a term has to be free, but here, we would want $y$ and $z$ to be bound. While the symbol $E$ was defined similarly with a comprehension, we need to show the existence and uniqueness of $E$ only once to introduce it with a definitional extension once and for all, as in \autoref{thm:extFOL}.

We represent $\lambda$-abstractions using a variable which is only valid under some \textit{context}, a set of formulas. 
For example, we can represent the term $\lambda x: \B. x$ using:
\begin{itemize}
    \item a \emph{variable} $\lambda_1$, along with the corresponding
    \item \emph{context}, the formula $\lambda_1 \in (\mathbb B \Rightarrow \mathbb B)\ \land\ \forall x\in \mathbb B.\ \lambda_1(x) = x$
\end{itemize} 
Formally, using the mechanism of extension by definition, we first extend FOST with constant and functional symbols for $\Rightarrow, \mathbb B, \mathbb N, E$, and $\app$, according to \autoref{def:setUniverse}. We then define $\embed{\cdot}$ as follows:

\begin{defin}[Embedding of HOL into FOST]
\label{defin:holInFost}
Reserve a special set of variables $\Lambda = \lbrace\lambda_1, \lambda_2, ...\rbrace$ that are used to represent lambda expressions and associate to every HOL term $t$ a single $i$. In practice, we use a global counter. We use the standard application notation for FOST Terms, so that, for example, $\lambda_2\, y\, y$ really means $\app(\app(\lambda_2, y), y)$.

$$
\begin{array}{lll}
    \embed{X} & = & X \\
    \embed{\Tl_1 \to \Tl_2} & = & \embed{\Tl_1}  \Rightarrow
    \embed{\Tl_1 } \\
    \embed{\mathcal B} & = & \mathbb B\\
    \embed{i} & = & \mathbb N \\
    \embed{x:A} & = & x \\
    \embed{=^\lambda : A \to A \to \mathcal B} & = & E(\embed{A}) \\ 
    \embed{(f: A \to B)(t: A) : B} & = & \embed{f: A \to B }\, \embed{t:B}\\
    \embed{(\lambda x: A.\ t:B): A \to B} & = & \lambda_i\, y_1\, ...\, y_n \\
     & & \text{Where $y_1$,...,$y_n$ are the free variables of $\lambda x. t$,}\\
     & & \text{and $\lambda_i$ is a variable symbol associated with the term $\lambda x.t$}
\end{array}
$$
\end{defin}
In the last line, $\lambda_i$ is a representation of the \textit{closure} of the lambda term $\lambda x.t$, and is intended to only be valid under the appropriate defining assumption.

There is another issue with this encoding, which is that we are losing type information associated to variables, as well as the HOL assumption that type variables cannot represent empty types. Fortunately, this can also be solved with contexts.

\subsection{HOL in FOST Using Contexts}
To translate propositions of HOL into FOST, we need to compute \textit{contexts} of HOL terms. We will need a \textit{non-emptiness context}, to handle type variables, a \textit{typing context}, to carry over information regarding types of variables, a \textit{definition context} to handle abstractions.

The following definition defines $\ctx^N$ (non-emptiness), $\ctx^T$ (variable typing) and $\ctx^D$ (definitions). Assume for simplicity and without loss of generality that variables typed differently have different identifiers, so that, for example $x$, $x:A$ and $x:B$ do not appear together in the same proof.

\begin{defin}[Non-Emptiness Context]
The typing context of an HOL term is the set of assumptions $A \neq \emptyset$ for every type variable in the term. This also includes type variables in the type signature of polymorphic constant symbols.
\end{defin}

\begin{defin}[Typing Context]
The typing context of an HOL term is a set of FOST formulas of the form $x \in T$ and is computed recursively as follows: 
$$
\begin{array}{lll}
    \ctx^T(x:T) & = & \lbrace x \in T \rbrace\\
    \ctx^T(c) & = & \emptyset \text{ for $c$ a constant symbol}\\
    \ctx^T(f\, t) & = & \ctx^T(f) \cup \ctx^T(t)\\
    \ctx^T(\lambda x:T.\, t) & = & \ctx^T(t) - \lbrace x \in T \rbrace
\end{array}
$$
\end{defin}

\begin{defin}[Definitional Context]
\label{defin:defContext}
The definition context of an HOL term is a set of FOST formulas whose free variables are from $\Lambda$ and from the set of type variables. It is computed as follows:
$$
\begin{array}{lll}
    \ctx^D(x:T) & = & \emptyset\\
    \ctx^D(c) & = & \emptyset\\
    \ctx^D(f\, t) & = & \ctx^D(t) \cup \ctx^D(t)\\
    \ctx^D(\lambda x:T. t) & = & 
        \ctx^D(t)\ \cup\ \\
    & & \lbrace(\lambda_i \in (\embed{T_1} {\Rightarrow}... {\Rightarrow}\embed{T_n} {\Rightarrow} \embed{T} {\Rightarrow} \embed{\type{(t)}}) \land \ \\
    & & \quad \forall y_1\in T_1,...,\forall y_n\in T_n. \lambda_i\, y_i\, ...\, y_n\, x = \embed{t}\rbrace \\
    & & \text{where $y_1:T_1, ..., y_n:T_n$ are the free variables of $t$ (without $x$)}.
\end{array}
$$ 
\end{defin}
$\lambda_i$ represents the closure of the $\lambda$-expression, as in the supercombinator compilation of functional programming languages \cite[Chapter 13]{DBLP:books/ph/Jones87}. Having $\lambda_i$ represent the closure of the lambda abstraction rather than the abstraction itself is necessary because otherwise the $y_i$'s would be free in the definition. But this should not be the case if they are supposed to be bound in an outer term. This is illustrated in the third formulas in the following \autoref{ex:contexts}.

The \textit{context}, $\ctx(t)$, of an HOL term $t$, is 
$\ctx^N(t) \cup \ctx^T(t) \cup \ctx^D(t)$.
\begin{ex}
\label{ex:contexts}
Let $x:X$, $y:Y$, $f:Y\to X$, $g:X\to Y$. We omit type annotations from lambda terms.
        $$ \begin{array}{lcl}
        
    \embed{\lambda x. x} &=& \lambda_1\\
    \ctx(\lambda x. x) &=& \lbrace X \neq \emptyset, \lambda_1 \in X{\Rightarrow}X \land \forall x\in X. (\lambda_1\, x) = x\rbrace\\
    \\

    \embed{(\lambda x. y)\,( f\, y)} &=& \lambda_2\, y\, (f\, y)\\
    \ctx((\lambda x. y)\, (f\, y)) &=& 
        \lbrace X \neq \emptyset, Y \neq \emptyset, y\in Y, g \in X\Rightarrow Y \\
    & & (\lambda_2 \in Y{\Rightarrow}X{\Rightarrow}Y )\land \forall y\in Y. \forall x\in X. (\lambda_2\, y\, x) = y\rbrace \\
    \\

    \embed{(\lambda y. (\lambda x.y) =^\lambda g)} &=& \lambda_4\, g\\
    \ctx((\lambda y. (\lambda x. y) =^\lambda g)\, y) &=& 
        \lbrace X \neq \emptyset, Y \neq \emptyset, f \in Y\Rightarrow X, \\ 
    & & (\lambda_2 \in Y{\Rightarrow}X{\Rightarrow}Y )\land \forall y\in Y. \forall x\in X. (\lambda_2\, y\, x) = y, \\
    & & (\lambda_4 \in (X{\Rightarrow}Y){\Rightarrow}Y{\Rightarrow}\mathbb B )\ \land \ \\
    & & 
        \forall g\in X\Rightarrow Y. \forall y\in Y. (\lambda_4\, g\, y = E(A)\, (\lambda_2\, y)\, g)
        \rbrace
    \\     
        \end{array} $$

Note that in the last example, the definition of $\lambda_4$ refers to $\lambda_2$, and binds the variable $y$ which is free in the $\lambda$-abstraction represented by $\lambda_2$. Without the closure, $y$ would be free in the definition of $\lambda_2$ and could not be bound in the definition of $\lambda_4$. Recall that $E(A)$ denotes the meaning of a (curried) equality relation on $A$.
\end{ex}

We can now define the embedding of sequents:
\begin{defin}
    Let $s = t_1,...,t_n \vdash t$ be an HOL sequent. Define the embedding $\embed s$ as
    $$
    \ctx{(t_1)},...,\ctx{(t_n)}, \ctx{(t)}, \embed{t_1 }= \top,...,\embed{t_n} = \top 
    \ \ \vdash\ \  \embed{t} =\top\text{.}
    $$
\end{defin}

\label{subsec:SyntacticEmbedding}

\subsection{Proof of Type Checking}
\label{subsec:typeChecking}
To produce proofs corresponding to type checking, we define a Lisa proof tactic \lstinline|ProofType(t: Term)| which for any term $t$ of type $T$ outputs a proof of
$$
\ctx(t) \vdash \embed{t} \in \embed{T}
$$
As abstractions are represented by typed variables applied to some other free variables in our encoding, the tactic only has to type applicative terms. For example, consider the term $(\lambda x:A.\, y:A)(z:A)$. The corresponding theorem of first-order set theory is:
$$
(\lambda_1 \in A \Rightarrow A \Rightarrow A) \land (\forall y\in A. \forall x\in A. \lambda_1\, x\, y = y), y \in A, z \in A\ \ \vdash\ \ (\lambda_1\, y)\, z \in A
$$
for which our approach generates a proof by recursing on the structure of $t$ and $T$, using the definition of function spaces.

\paragraph*{Polymorphism}
More interesting is the typing of polymorphic constants such as HOL equality $=^\lambda$. Its HOL type is $A \to A \to \B$ and its interpretation according to \autoref{defin:holInFost} is $E(A)$. Hence, the corresponding typing judgement proven by \lstinline|ProofType| should be $E(A) \in A \to A \to \B$. 
In simply typed lambda calculus with explicit polymorphism (like System F, see for example \cite{barendregtLambdaCalculusTypes2013}), $=^\lambda$ would be given the type $\Lambda A. A \to A \to \B$ to $E$. The corresponding property for $E$ in FOST is $\forall A. E(A) \in A \Rightarrow A \Rightarrow \mathbb B$. This is easy to represent in our translation using free set variables in sequents.
We added support for such top-level polymorphism to the \lstinline|ProofType| tactic, so that it can automatically type polymorphic constants embedded this way. 


\subsection{Simulating HOL Proofs}
The goal of the section is to demonstrate that HOL Light proof steps can be simulated by proofs in our encoding.

\begin{thm}[Simulating HOL Proofs in FOST]
    \label{thm:ruleCorrectness}
    Let 
    \begin{center}
        \AxiomC{$s_1$}
        \AxiomC{$...$}
        \AxiomC{$s_n$}
        \TrinaryInfC{$s$}
        \DisplayProof
    \end{center}
    be an instance of a deduction rule of HOL from \autoref{fig:rulesHOL}. Then 
    \begin{center}
        \AxiomC{$\embed{s_1}$}
        \AxiomC{$...$}
        \AxiomC{$\embed{s_n}$}
        \TrinaryInfC{$\embed{s}$}
        \DisplayProof
    \end{center}
    is admissible in FOST (rules of sequent calculus and axioms of set theory).
\end{thm}

We state three auxiliary theorems of FOST which will be necessary for the simulation:

\newcommand{\siff}{\mathop{\Leftrightarrow}}
\begin{lem}
\label{lem:collect}
The following statements are theorems of FOST:
\begin{align*}
    x \in A, y \in A\ \vdash\ (E(A)\, x\, y = \top) \siff (x = y)                              \label{lem:Ecorrect}\tag{Correctness of $E$} \\
    f \in A\Rightarrow B, g \in A\Rightarrow B, \forall x \in A. f\, x = g\, x \ \vdash\ f = g \label{lem:funExt}\tag{Functional Extensionality}\\
    p \in \mathbb B, q \in \mathbb B, (p = \top) \siff (q = \top)\ \vdash\ p = q               \label{lem:propExt}\tag{Propositional Extensionality} 
\end{align*}
\end{lem}
The simulation of a proof step can in general be split in two parts: first produce a proof under arbitrary typing and context assumptions, and then handle the modifications in context. For example, let $x:\B, f:\B \to \B, g:\B \to \B$ and consider a \TRANS step deducing
    \begin{center}
        \AxiomC{$\Gamma \vdash x =^\lambda f\, x$}
        \AxiomC{$\Gamma \vdash f\, x =^\lambda g\, x$}
        \BinaryInfC{$\Gamma \vdash x =^\lambda g\, x$}
        \DisplayProof
    \end{center}
and let $c_\Gamma = \ctx(\Gamma)$.
We wish to obtain a proof of
    \begin{center}
        \AxiomC{$x \in \mathbb B, f \in \mathbb B{\Rightarrow}\mathbb B,                                         c_\Gamma, \embed{\Gamma} \vdash (x =^\lambda f\, x) = \top$}
        \AxiomC{$x \in \mathbb B, f \in \mathbb B{\Rightarrow}\mathbb B, g \in \mathbb B{\Rightarrow}\mathbb B,  c_\Gamma, \embed{\Gamma} \vdash (f\, x =^\lambda g\, x) = \top$}
        \BinaryInfC{$x \in \mathbb B, g \in \mathbb B{\Rightarrow}\mathbb B,                                     c_\Gamma, \embed{\Gamma} \vdash (x =^\lambda g\, x) = \top$}
        \DisplayProof
    \end{center}
    This should follow from applying \ref{lem:Ecorrect} (\autoref{lem:collect}) to each premise, using transitivity of first order equality, and applying back \ref{lem:Ecorrect} (\autoref{lem:collect})  to the result. However, to apply this lemma, we need the facts $f\, x \in \mathbb B$ and $g\, x \in \mathbb B$. Moreover, the $f \in \mathbb B{\Rightarrow}\mathbb B$ assumption from the premise will stay in the conclusion, yielding:
    $$
    x \in \mathbb B, f \in \mathbb B{\Rightarrow}\mathbb B, g \in \mathbb B{\Rightarrow}\mathbb B, f\, x \in \mathbb B, g\, x \in \mathbb B, 
    c_\Gamma, \embed{\Gamma} \vdash (x =^\lambda g\, x) = \top
    $$
    which is a correct conclusion but contains too many assumptions. Fortunately, these assumptions can be eliminated.

\paragraph*{Eliminating lingering assumptions}

    Pursuing the example above, let $L = \lbrace x \in \mathbb B, g \in \mathbb B{\Rightarrow}\mathbb B, c_\Gamma, \embed{\Gamma}\rbrace$ and $R = (x =^\lambda g\, x) = \top$. We want to simulate the following proof step:
    \begin{center}
        \AxiomC{$f \in \mathbb B{\Rightarrow}\mathbb B, f\, x \in \mathbb B, g\, x \in \mathbb B, L \vdash R$}
        \UnaryInfC{$L \vdash R$}
        \DisplayProof
    \end{center}
    First, we prove (automatically) the non-elementary typing assumptions $(x \in \mathbb B, f \in \mathbb B{\Rightarrow}\mathbb B) \vdash f\, x \in \mathbb B$ by induction over the structure of $f\, x$ (as in \autoref{subsec:typeChecking}) and similarly for g.
    Then, note that $f$ is free everywhere but in its typing assumption: we can quantify it to $\exists f. f \in \mathbb B {\Rightarrow} \mathbb B$ using the LeftExists rule from first order logic. Now this statement is provable: It can be deduced from the non-emptiness of $\mathbb B$. Formally, we obtain the following proof:
    \begin{center}
        \AxiomC{$f \in \mathbb B{\Rightarrow}\mathbb B, f\, x \in \mathbb B, g\, x \in \mathbb B, L \vdash R$}
        \AxiomC{...}
        \UnaryInfC{$x \in \mathbb B, f \in \mathbb B{\Rightarrow}\mathbb B \vdash f\, x \in \mathbb B$}
        \RightLabel{ Cut\kern-10em}
        \BinaryInfC{$f \in \mathbb B{\Rightarrow}\mathbb B, g\, x \in \mathbb B, L \vdash R$}
        \UnaryInfC{$\vdots$}
        \noLine\UnaryInfC{}
        \UnaryInfC{$f \in \mathbb B{\Rightarrow}\mathbb B, L \vdash R$}
        \RightLabel{ LeftExists\kern-10em}\UnaryInfC{$\exists f. f \in \mathbb B{\Rightarrow}\mathbb B, L \vdash R$}
        \AxiomC{$...$}
        \UnaryInfC{$\vdash \exists f. f \in \mathbb B{\Rightarrow}\mathbb B$}
        \RightLabel{ Cut}\BinaryInfC{$L \vdash R$}
        \DisplayProof
    \end{center}

    This example covers statements corresponding to type judgement and typing context.
    In general, there are three kinds of context formulas we need to eliminate: lambda definitions, variable type assignment, type variables non-emptiness. We implement a proof tactic, called \CLEAN{} which eliminates context formulas iteratively:
    \begin{enumerate}
        \item Find in the context a definition $\mathit{def}(\lambda_i)$ such that $\lambda_i$ does not appear anywhere else.
            Then, using LeftExists, generalize the left handside to $\exists \lambda_i. \mathit{def}(\lambda_i)$.
            Prove $\exists \lambda_i. \mathit{def}(\lambda_i)$. This is always possible using the adequate type-nonemptiness assumption.
            Eliminate $\exists \lambda_i. \mathit{def}(\lambda_i)$ using the Cut rule. Iterate on the next definition.
        \item Find a variable type assignment $x\in T$. Using LeftExists, generalize to $\exists x. x\in T$.
            Using the type variables non-Emptiness assumptions, prove that $\exists x. x \in T$ (i.e. $T$ is non-empty).
            Eliminate $\exists x. x \in T$. Iterate on next unused variable.
        \item Find a non-Emptiness assumption $A \neq \emptyset$ for a type variable that does not appear anywhere else. 
            Using LeftExists, generalize to $\exists A. A \neq \emptyset$, which is of course provable without assumption, and eliminate it.
            Iterate on the next unused type variable.
    \end{enumerate}
    We make every tactic that possibly eliminates subterms (that is, \TRANS, \ABS, \EQMP, \INST{} and \INSTTYPE{}) call \CLEAN to eliminate lingering assumptions.

\paragraph*{Simulating HOL steps}
    We briefly hint at how steps of HOL can be simulated in FOST, leaving implicit concerns regarding proofs of type checking and context elimination, which were addressed above.
\begin{itemize}
    \item \REFL is simulated with \ref{lem:Ecorrect} and reflexivity of first-order equality
    \item \TRANS is similarly simulated with \ref{lem:Ecorrect} and reflexivity of first order equality. Note that in HOL Light, \TRANS requires only alpha-equivalence of the shared terms of the two premises. We explain how this can be handled without assuming that all alpha-equivalent expressions are represented by the same $\lambda_i$ in the next paragraph.
    \item \MKCOMB is simulated with from \ref{lem:Ecorrect} and substitution of equals for equals of first-order logic.
    \item \ABS and \ETA follow from the definition of the $\lambda_i$ and from \ref{lem:funExt} (\autoref{lem:collect}).
    \item \BETA steps are proven directly from the definition of the $\lambda_i$.
    \item \ASSUME is simply a Hypothesis step in Sequent calculus
    \item \EQMP is simulated with \ref{lem:Ecorrect} and substitution of equals. Similar \TRANS, it is subsequently made to support alpha-equivalence. 
    \item \DEDUCT steps are proven with \ref{lem:propExt}
    \item \INST follows from instantiation of free variables in first order logic, except that doing so changes the shape of embeddings of abstractions to a non-canonical representation, which need to be transformed back into a canonical representation. We explain this mechanism in detail in the following paragraph.
    \item \INSTTYPE is simply instantiation of free variables.
\end{itemize}

\paragraph*{Alpha Equivalence}
The steps \TRANS and \EQMP each take 2 premises with the added requirement that they share some subterm. For concreteness, consider the \TRANS step:

\begin{center}
    \AxiomC{$\Gamma \vdashhol s =_A t$}
    \AxiomC{$\Delta \vdashhol t =_A u$}
    \RightLabel{\text { \TRANS}}
    \BinaryInfC{$\Gamma, \Delta \vdashhol s =_A u$}
    \DisplayProof
\end{center}

In HOL Light the two terms $t_1$ and $t_2$ in the premises are required to be identical \textit{up to alpha equivalence}. However, alpha equivalence does not naturally hold in our encoding: two alpha-equivalent abstractions may be represented by different variables $w_i$ from \autoref{defin:holInFost}.
(In fact, in the absence of memoization, even two occurrences of the same lambda can be represented by different variables. In practice, for our import from HOL Light we perform memoization in the constructor of abstractions $\lambda$\lstinline{(x:Var, body: Term)} using de Brujin indices so that alpha equivalent terms HOL terms are mapped to the exact same FOST expression FOST for efficiency. That said, we still wish to show how to support alpha equivalence as a rule.) 

For example, consider symbols $\lambda_1$ and $\lambda_2$, representing abstractions, with the definitions:
$$
(\lambda_1 \in A \Rightarrow A) \land (\forall x\in A. \lambda_1 \, x = x)
$$
$$
(\lambda_2 \in A \Rightarrow A) \land (\forall y\in A. \lambda_2 \, y = y)
$$
Here, we can use the fact that our local definitions of lambda terms ensure not only existence, but also uniqueness. In particular, under those two assumptions, $\lambda_1 = \lambda_2$ is a consequence of the extensionality of set-theoretic functions.
In fact, using the Eta axiom from HOL (implemented instead as a deduction step \ETA), alpha equivalence is provable and does not need to be assumed.

\begin{defin}[Tactic for Alpha-Conversion]
Let \uTRANS and \uEQMP be restrictions of \TRANS and \EQMP not supporting alpha-equivalence. We implement a tactic 
\begin{center}
    \AxiomC{}
    \RightLabel{\ALPHACONV}
    \UnaryInfC{$\vdashhol \lambda x.t = \lambda y.t[x:=y]$}
    \DisplayProof
\end{center}
where $$\text{\ALPHACONV}\, x\, y\, t = \text{\uTRANS}\, (\text{\ETA}\, y\, (\lambda x.\, t))\, (\text{\ABS}\, (\text{\INST}\, (\text{\BETA}\, x\, t)\, x\, y)\, y)$$
Note that the first argument of \uTRANS proves $\lambda x. t = \lambda y. (\lambda x.t) y$ and the second proves $\lambda y. (\lambda x.t) y = \lambda y.t[x:=y]$

We can then define a tactic proving the following:
\begin{center}
    \AxiomC{}
    \RightLabel{\ALPHAEQUIVALENCE \text{ (if $t$ and $u$ are alpha-equivalent)}}
    \UnaryInfC{$\vdashhol t = u$}
    \DisplayProof
\end{center}
which proves the equality by applying \ALPHACONV recursively on $t$ and $u$.
Finally, we can define the complete versions \TRANS{} and \EQMP{}, which apply \ALPHAEQUIVALENCE{} if the shared terms in the input are not strictly equal.
\end{defin}

\paragraph*{INST Proof Step and Normalization}
It may seem at first glance that \INST{} is a very easily simulated step: first-order logic admits instantiation of free variables across a sequent (in fact, LISA offers this as a built-in proof step). This however fails to preserve the structure of the embedding. For concreteness again, let $x:A, y:A, p:\mathcal{B}$ and consider the following simple provable HOL statement and its embedding in FOST:
$$
\begin{array}{r c l}
S &=& \vdash (\lambda x. p = p) y \\
\embed{S} &=& p\in \B, y \in A, \mathit{def}_{\lambda_1} \vdash (\lambda_1\, p\, y) = \top
\end{array}
$$
where $\mathit{def}_{\lambda_1} = (\lambda_1 \in B \Rightarrow A \Rightarrow \B) \land (\forall p\in B. \forall x\in A. \lambda_1\, p\, x = (p=^\lambda p))$.
Now, suppose $f:: B \Rightarrow B$ is also a variable and consider the effect of the instantiation $p := (f p)$ 
$$
\begin{array}{r c l}
S_{[p := (f p)]} &=& \vdash (\lambda x. (f p) = (f p)) y \\
\embed{S}_{[p := (f p)]} &=& p\in \B, y \in A, \mathit{def}_{\lambda_1} \vdash (\lambda_1\, (f p)\, y) = \top
\end{array}
$$
But on the other hand, we have
$$
\begin{array}{r c l}
\embed{S_{[p := (f p)]}} &=& p\in \B, y \in A, \mathit{def}_{\lambda_2} \vdash (\lambda_1\, f\, p\, y) = \top
\end{array}, \mbox{ where }
$$
$\mathit{def}_{\lambda_2} = (\lambda_2 \in (B{\Rightarrow} B) {\Rightarrow} B {\Rightarrow} A {\Rightarrow} \B) \land
    (\forall f\in (B {\Rightarrow} B). \forall p\in B. \forall x\in A. \lambda_2\, f\, p\, x = (p=^\lambda p))$.
So, instantiation and embedding do not commute. Moreover, the shape of $S_{[p := (f p)]}$ does not correspond to the canonical specification of the embedding of HOL terms described in \autoref{defin:holInFost}.
\begin{defin}
    (The embedding of) an abstraction term $t$ is in \textit{closure-canonical} form if it is of the form $\lambda_i\, x\, y\, z...$ where $x, y, z...$ are the free variables of $t$ and $\lambda_i$ is a symbol whose local context is as defined by \autoref{defin:defContext}. A term is in closure-canonical form if all its subterms are in closure-canonical form.
\end{defin}
$\lambda_2\, f\, p\, x$ is in closure-canonical form, as any term produced by \autoref{defin:holInFost}, but $\embed{S_{[p := (f p)]}}$ is not, because  the subterm $\lambda_1\, (f p)\, y$ is not in canonical form.
Hence, even though it denotes an equivalent statement, $S_{[p := (f p)]}$ is not a legal expression whose shape other proof tactics expect to receive.
To solve this, we implemented a tactic that recursively transforms any non-canonical representation of an HOL term into its closure-canonical form and prove equality between the two. This then allows the \INST tactic to output a statement in canonical form.

This concludes our simulation of the various proof tactics in FOST leading to \autoref{thm:ruleCorrectness}. 
\begin{cor}
    \label{cor:seqCorrectness}
    Let $s$ be an HOL sequent. Then a proof of $s$ in HOL can be transformed in a proof of $\embed{s}$ in FOST.
\end{cor}
We have implemented the transformation and the above tactics in Lisa.

\subsection{Defining new constants}
\label{subsec:constantdef}

HOL Light allows the introduction of new definitions which serve as shorthand for existing terms. This is also possible in Lisa, where if we produce a theorem of the form $\exists! x. P(x)$, we obtain a new constant $c$ and the property $\forall x. P(x) \iff x = c$. However, for this extension to be sound, the definition of a constant can not contain free variables, as otherwise defining $c := x$ would allow proving $\forall x. c = x$. For definitions from HOL, consider for example the term defining the universal quantifier $!$ in HOL Light (\lstinline|bool.ml: 243|):
$$
 \lambda P:A\to\B.P = \lambda x:A. \top
$$
It is represented as a certain variable $\lambda_2$ while the subterm $\lambda x:A. \top$ is represented by some symbol $\lambda_1$. As in the elimination of contexts, we can prove $\exists! \lambda_2. \ctx{(\lambda_2)}$, which matches the requirement of extension by definition. The type variable $A$ is reflected as an explicit parameter of the constant (which means that $!$ is an applied function symbol $!(A)$).
However, $\lambda_1$ will be free in $\ctx{(\lambda_2)}$, so we need to bundle with the definition of $\lambda_2$ all the context definitions that it refers to and prove existence and uniqueness

It is easy to prove, that such a symbol exists under the assumption of the usual context:
$$
\ctx{(\lambda_1)}, \ctx{(\lambda_2)} \vdash \exists! c.\; c = \lambda_2 
$$
However when quantifying all assumptions, this will only yield
$$
\forall \lambda_1.\; \ctx{(\lambda_1)} \implies \forall \lambda_2.\; \ctx{(\lambda_2)} \implies \exists! c.\; c = \lambda_2 
$$
while the mechanism of extension by definition requires the $\exists !$ quantifier to be first in the definition. We use an additional FOL theorem that allows us to swap the universal and unique-existential quantifiers 
$$\exists! x. P(x) \implies ((\forall x.\; P(x) \implies \exists! y.\; Q(x, y)) \iff (\exists! y.\; \forall x. P(x) \implies Q(x, y)))$$
This fact, alongside proofs that the terms $\lambda_i$ are uniquely defined by their contexts, we can swap the quantifiers one-by-one to produce the final justification for the definition:
$$
\exists! c.\;\forall \lambda_1.\; \ctx{(\lambda_1)} \implies \forall \lambda_2.\;   \ctx{(\lambda_2)} \implies  c = \lambda_2 ~.
$$

We generate this proof automatically in our implementation. The proof of typing is generated alongside the symbol by type checking the definition. The theorem corresponding to the definition under appropriate context
$\ctx{(\lambda_1)}, \ctx{(\lambda_2)} ~\vdash~ !(A) = \lambda_2$
is also generated automaticaly.

%% file: ADT.tex
We want to bring the benefits of types into FOST. In particular, algebraic data types are useful when reasoning about inductive data structures such as lists or trees. Their encoding is generally hidden to the user, who only obtains access to their characteristic theorems and definitional mechanisms. We want Lisa to incorporate such mechanisms. Even though ADTs can be encoded within HOL \cite{Melham1989}, we choose instead for Lisa's implementation to directly define them in FOST. We therefore avoid going through an intermediate encoding but also lay the foundations of further generalization.
We start by giving a syntactic definition of algebraic data types.

\begin{defin}[Algebraic data types]
    An \textit{algebraic data type} in set theory is a set $A$ equipped with a finite set of functions $c_i: T^i_1 \Rightarrow \dots \Rightarrow T^i_n \Rightarrow A$ referred to as constructors. All elements of $A$ have to be in the image of one of exactly one of its constructors. $T^i_j$ can refer  to $A$ itself, giving algebraic data types their recursive behavior.
\end{defin}

 We want to allow defining and reasoning about ADTs directly in FOST. Specifically, consider an ADT specification $\{c_i: (S^i_1, \dots, S^i_n)\}_{i \leq m}$, where $S^i_j$ is either a term with no variables, or a special symbol referring to the ADT itself. We want to output a set $A$ and functions $\{c_i\}_{i \leq m}$ with the following properties:

\begin{itemize}
    \item (Typing) For all $i \leq m$, $c_i \in (S^i_1 \Rightarrow \dots \Rightarrow S^i_n \Rightarrow A)$ 
    \item (Injectivity 1) Every $c_i$ is injective.
    \item (Injectivity 2) For $x, y\in A$, if $x \in \text{Im}(c_i)$, $y \in \text{Im}(c_j)$ and $i \neq j$ then $x \neq y$
    \item (Structural induction) $A$ is the smallest set closed under the constructors $c_i$'s. This allows us to write proofs by induction on the structure of the ADT.
\end{itemize}

\begin{ex}
    Consider the type of boolean linked lists, with specification 
    \begin{center}
        $\text{listbool}=\{\text{nil}: (), \text{cons}: (\mathbb{B}, \text{listbool})\}$
    \end{center}
    We want to generate a set listbool and constructors nil and cons such that nil $\in$ listbool and cons $\in \mathbb{B}\Rightarrow \text{listbool} \Rightarrow \text{listbool}$. The above properties should hold; for instance cons $\top$ nil $\neq$ cons $\bot$ nil.
\end{ex}

We next present our formalization as implemented in Lisa using set theoretic axiomatization. We represent an algebraic data type as a set $A$ of tuples containing the tag of the constructor and the arguments given to it. This ensures that elements of $A$ are in the image of exactly one constructor.
For the listbool example, cons $\top$ nil is hence represented as the tuple (\consttag$_{\text{cons}}$, $\top$, (\consttag$_{\text{nil}}$)). In this setting, tags are arbitrary terms that differentiate constructors. They can, for example, be natural numbers or some encoding of the name of the constructor. 
We define the set $A$ as the least fixpoint of the function
\begin{center}
    $F(H) = \displaystyle\bigcup\limits_{i \leq m} \left \lbrace (\consttag_{c_i}, x_1, \dots, x_n)\ \bigg|\ x_k \in \begin{cases}
        H &\text{if $S^i_k$ is a self-reference}\\
        S^i_k & \text{otherwise} 
    \end{cases}
    \right\rbrace$
\end{center}
The existence of $F(S)$ for every set $S$ is guaranteed by the replacement and the union axioms. 
In order to characterize the least fixpoint of $F$, we use the recursion theorem schema of ZF to obtain a unique function $f$ with domain $\mathbb N$ (which is also the smallest infinite ordinal $\omega$) such that
$$f(0)     = \emptyset $$
$$f(n + 1) = F(f(n))$$

Intuitively, $f(n)$ corresponds to all instances of $A$ of height smaller or equal than $n$. For example, for list of booleans, $f(2)$ is the set 
$$\left\lbrace(\consttag_{\text{nil}}), (\consttag_{\text{cons}}, \top, (\consttag_{\text{nil}})), (\consttag_{\text{cons}}, \bot, (\consttag_{\text{nil}}))\right\rbrace$$

\begin{lem}
    The class function $F$ admits a least fixpoint given by $A := \bigcup_{n \in \omega} f(n)$.
\end{lem}
\begin{proof}
    If $x_k \in A$ then there is a $n_k \in \omega$ such that $x_k \in f(n_k)$. Since $F$ is monotonic $x_k \in f( \max_{x_k \in A}\, n_k)$. Therefore, for every $(\consttag_{c_i}, x_1, \dots, x_n) \in F(A)$, we have $(\consttag_{c_i}, x_1, \dots, x_n) \in f( \max_{x_k \in A}\, n_k + 1) \subseteq A$.
\end{proof}

\begin{defin}
     We define $c_i$ as the function in $ S^i_1 \Rightarrow \dots \Rightarrow S^i_n \Rightarrow A$ such that
    \begin{center}
        $c_i\, x_1\, \dots\, x_n = (\consttag_{c_i}, x_1, \dots, x_n)$
    \end{center}
\end{defin}

Now that we have a formal definition of $A$ and $c_i$, we prove that they fulfill the above properties.

\begin{thm} 
\label{thm:adtStatements}
Let $A$ and $\{c_i\}_{i \leq m}$ constructed as above. The following statements hold in FOST
    \begin{align*}
    x_1 \in S^i_1, \dots, x_n \in S^i_n \vdash c_i\, x_1\, \dots\, x_n \in A                           \tag{Typing} \\
    x_k \in S^i_k,\, y_k \in S^i_k,\, c_i \, x_1\, \dots\, x_n = c_i \,y_1\, \dots\, y_n \vdash \bigwedge_{k \leq n} x_k = y_k \tag{Injectivity 1} \\
    x_k \in S^i_k,\, y_k \in S^j_k,\, i \neq j \vdash c_i \,x_1 \, \dots \, x_n \neq c_j \, y_1 \, \dots \, y_{n'} \tag{Injectivity 2}
\end{align*}
\end{thm}
\begin{proof}
    For typing, we have $c_i\, x_1\, \dots\, x_n = (\consttag_{c_i}, x_1, \dots, x_n) \in F(A) = A$.\\
    We know that tuples are injective, that is 
    \begin{center}
        $\vdash (\consttag_{c_i}, x_1, \dots, x_n) = (\consttag_{c_j}, y_1, \dots, y_n) \iff \bigwedge_{k \leq n} x_k = y_k \land \consttag_{c_i} = \consttag_{c_j}$
    \end{center}
    Moreover, tuples of different arities are not equal.
    Injectivity 1 follows from the right implication while Injectivity 2 from the left one and the fact that tags are uniquely assigned to constructors.
\end{proof}
\begin{thm}
    \label{thm:structuralInduction}
    Structural induction schema holds on $A$.
    \begin{center}
        $\displaystyle \bigwedge_{i \leq m} \left(\forall x^i_1 \in S^i_1.\  \hat{P}(x^i_1) \Longrightarrow \dots \Longrightarrow \forall x^i_{n} \in S^i_{n}. \ \hat{P}(x^i_n) \Longrightarrow   P(c_i \, x^i_1\, \dots \, x^i_n)\right) \vdash \forall x  \in A.\ P(x)$
    \end{center}
    where $\hat{P}(x^i_k) = \begin{cases}
        P(x^i_k) &\text{if $S^i_k = A$} \\
        \top &\text{if $S^i_k \neq A$}
    \end{cases}$
    
\end{thm}
\begin{proof}
    We first show that for every $n \in \mathbb{N}$, the theorem holds when replacing $A$ by $f(n)$. This follows by induction on $n$. Since by definition of $A$, every $x \in A$ is in $f(n)$ for some $n$, the statement holds for every element of $A$. 
\end{proof}

\subsubsection*{Polymorphic algebraic data types}

Algebraic data types can be polymorphic, meaning that the specification of the constructors contain type parameters. This allows, for example, reasoning over generic lists instead of lists of a specific type. We extend our mechanization of ADTs to support such polymorphism. To do so, we generalize  $A$ and $\{c_i\}$ to be class functions instead of constant symbols.

Formally, let  $\{c_i: (S^i_1, \dots, S^i_n)\}_{i \leq m}$ be the specification of an algebraic data type that possibly contains variable symbols $X_1, \dots, X_l$. We define $A(X_1, \dots, X_l)$ as the fixpoint of $F(X_1, \dots, X_l)$ where the construction of $F$ carries the same behavior as above.
Using this definition, all the properties about ADT are preserved and the constructions is essentially the same. 

As in \autoref{subsec:typeChecking}, we give a top level polymorphic type to the function symbols $A$ and $c_i$, so that they can similarly be typechecked. This also implies that all the tactics from the previous section are compatible with terms referring some ADT $A$ and its constructors. To conclude, we show an example of polymorphic lists in Lisa.

\begin{ex}
    The user can define polymorphic lists with the following syntax:
    \begin{lstlisting}[language=scala]
    val define(list: ADT, constructors(nil, cons)) = () | (T, list)
    \end{lstlisting}
    where \lstinline|list|, \lstinline|nil|, and \lstinline|cons| are new function symbols. \lstinline|list| is such that \lstinline|list(T)| is a set containing all lists over \lstinline|T| \lstinline|nil|is a constructor taking one type parameter and no argument and \lstinline|cons(T)| is a constructor taking one type parameter \lstinline|T| as well as an element of type \lstinline|T| and an element of \lstinline|list(T)| as arguments. This declaration also automatically proves \cref{thm:adtStatements} and \cref{thm:structuralInduction}
    Our typing tactics can use these properties to type check any expressions containing list, nil and cons.

\end{ex}

%% file: hol-import.tex

While the embedding of HOL as described above allows writing HOL proofs directly in Lisa, we also implement a prototype to attempt the automatic import of theorems and definitions from HOL Light.
We chose HOL Light for our import due to its simple foundations, its large library and easily accessible proof export. With some additional work in matching proof steps, the same method can be adapted to other members of the HOL family.

Since the HOL Light kernel does not keep track of proof objects by default, we
rely on the \lstinline|ProofTrace| export system packaged in the HOL Light repository \cite{polu2019prooftrace}.
The system provides a patch to the HOL Light kernel to track every proven
statement in the system's execution. The stored proofs are finally exported to
JSON files. We modified the existing JSON output syntax slightly to allow its
automatic import using standard JSON libraries for Scala. The terms are exported
as strings with a simple and unambiguous grammar, and are parsed back by Lisa. 
After reading the JSON files, the proofs (with indexed steps) are transformed
into proof DAGs in an intermediate representation, and finally transformed to
Lisa theorems. 

Given the Lisa tactics we developed for this purpose, the translation of HOL Light theorems is straightforward, and proceeds by recursing on the proof DAG obtained from HOL Light, translating each proof step to the equivalent Lisa tactic call. Although constant definitions also appear as a single proof step, \lstinline|DEFINITION|, they must be dealt with separately, as in~\autoref{subsec:constantdef}.

After defining a polymorphic constant, however, we change its signature compared to the HOL Light version. For example, the universal quantifier \lstinline|!: (A -> bool) -> bool| becomes a class function \lstinline|!| such that $\forall A.$ \lstinline|!(A)| $\in (\text{\lstinline|A|} \Rightarrow \B) \Rightarrow \B$. On subsequent occurrences of \lstinline|!| in the import, it occurs with an ascribed type, say \lstinline|!: (T -> bool) -> bool|. This instantiated type is matched against the original, polymorphic type to find the substitution \lstinline|A| $\mapsto$ \lstinline|T|. The definition is correspondingly instantiated, and the occurrence is replaced by the Lisa term \lstinline|!(T)|.

Our prototype implementation of the embedding described here produces a large overhead during this import. We process the first 15 named definitions and theorems as defined in the HOL Light library, producing 1716 HOL Light kernel steps. These are expanded to just over 300,000 Lisa kernel steps, with the reconstruction and verification taking 93 seconds on a laptop running Linux with an i7-1165G7 CPU and 16GB RAM.